\documentclass[[prb,twocolumn,superscriptaddress]{revtex4}

\usepackage{graphicx}
\usepackage{dcolumn}
\usepackage{bm}
\usepackage{textcomp}
\def\cm{cm$^{-1}$}

\begin{document}
\title{The quest for electronic ferroelectricity in organic charge-transfer crystals}
\author{Alberto Girlando}
\affiliation{Group on Molecular Materials for Advanced Applications (MMAA), c/o  Dept. of Chemistry, Life Sciences and Environmental Sustainability, Parco Area delle Scienze 17, 43124 Parma, Italy}

\begin{abstract}
Organic ferroelectric materials are in demand in the growing field of environmentally friendly, lightweight electronics. Donor-Acceptor charge transfer crystals have been recently proposed as a new class of organic ferroelectrics, which may possess a new kind of ferroelectricity, the so-called electronic ferroelectricity, larger and with faster polarity switching in comparison with conventional, inorganic or organic, ferroelectrics. The current research aimed at achieving ambient conditions electronic ferroelectricity in organic charge transfer crystals is shortly reviewed, in such a way to evidence the emerging criteria that have to be fulfilled to reach this challenging goal.  
\end{abstract}

\maketitle

\section{Introduction}
Ferroelectricity, i.e., spontaneous electrical polarization, is a phenomenon analogous to ferromagnetism, yet its discovery in actual materials is relatively recent, dating back to the last century \cite{lines_book77}. On the other hand, in the emerging field of lightweight and environmentally friendly materials for electronics, the quest for organic ferroelectrics operating at room temperature has proven to be more promising than that of ferromagnetic solids \cite{horiuchi_08}. The polar nature of ferroelectricity of course requires that the crystal structure does not have an inversion center, and in general the ferroelectrics are divided into different classes on the basis on how this symmetry loss is obtained. In recent years, attention has focused on organic mixed stack charge-transfer (ms-CT) co-crystals, made up by planar $\pi$-electron donor (D) and acceptor (A) molecules alternating along the stack direction. These materials are characterized by $\rho$, the degree of CT, ranging from 0 to 1, with $\rho \sim$ 0.5 separating the neutral (N) from the ionic (I) ground state \cite{masino_17,davino_17a}. Increase of the Madelung energy, following lattice contraction by lowering temperature, may induce a peculiar phase transition, the N to I one, with $\rho$ crossing the N-I borderline. Ionic systems are subject to the Peierls instability, yielding to dimerization of the stack, hence satisfying the above mentioned criterion for ferroelectricity. In addition, in ms-CT crystals the ferroelectricity may occur through a new mechanism, involving the $\pi$-electronic cloud along the stack, yielding high values of polarization for an organic system. In this paper, the quest for “electronic ferroelectricity” in ms-CT crystals will be shortly reviewed.

\section{Basic characteristics and classification of ferroelectrics}
\begin{figure}
	\centering
	\includegraphics[width=1.\linewidth]{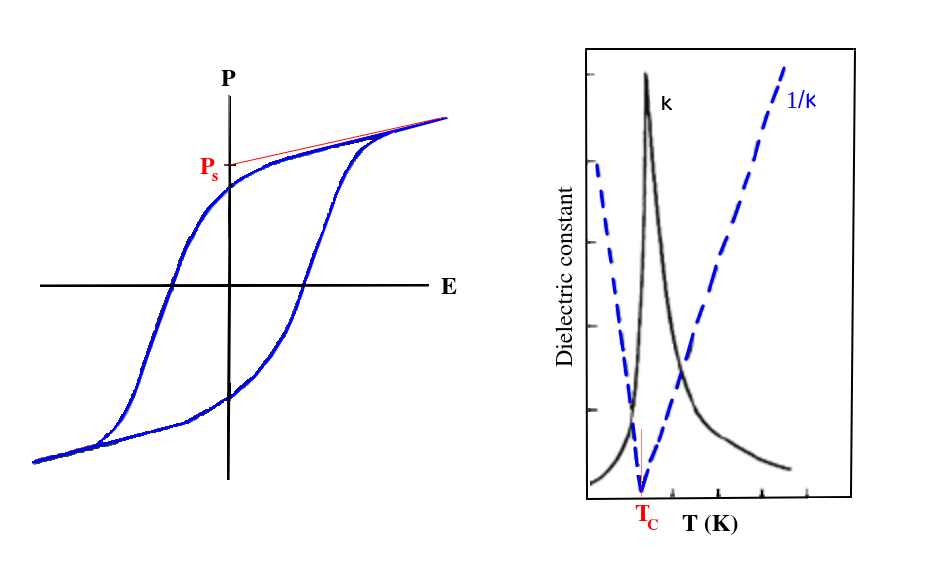}
	\caption{Ferroelectrics: Typical hysteresis loop ($\boldsymbol{P}$ vs. $\boldsymbol{E}$) and temperature dependence of the dielectric constant $\kappa$.}
	\label{fig:fig1ferro}
\end{figure}

A crystal possessing spontaneous electrical polarization, i.e., electric dipole, in general is not able to orient nearby electric dipoles since the polarization is canceled out by the so-called leakage currents, due to electrons moving through the crystal, ions moving through the air etc.  A net dipole can be temporarily generated by changing temperature, which changes the polarization due to a change in the equilibrium position of the atoms in the crystal. We then have the pyroelectric effect, which lasts until the moving electrons adapt to the new equilibrium. A polar, pyroelectric crystal is not necessarily a ferroelectric crystal, because ferroelectricity implies that the direction of the dipoles can be changed by an external electrical field, giving rise to the typical hysteresis loop of the polarization $\boldsymbol{P}$ vs. electric field $\boldsymbol{E}$ shown on the left side of Figure \ref{fig:fig1ferro}. The ferroelectric loop indicates the alignment of the electric dipole domains inside the crystal, and also implies that above a certain temperature, called Curie temperature $T_c$, there is a phase transition to a paraelectric, non-polar (centrosymmetric) phase. 

\begin{table*}[t]
\caption{Basic parameters of selected ferroelectrics}
\begin{tabular}{|l|c|c|c|cc|}
		\hline 
		~Material$^a$ & ~Curie temp. $T_c$ (K)~ & ~$\boldsymbol{P}_s$ ($\mu$C cm$^{-2}$ @ $T_c$)~ & Max. diel. const. $\kappa_\mathrm{max}$ &&  Ref. \\ 
		\hline 
~Rochelle salt & 297 & 0.25 @ 276 & 4 $\cdot 10^3$ && \cite{shiozaki_book_06} \\ 
~BaTiO$_3$ & 381  & 26 @ 300 & $10^4$ && \cite{shiozaki_book_06} \\  
~Thiourea & 169 & 3.2 @ 120 & $10^4$ && \cite{goldsmith_59} \\ 
~VDF$_{0.65}$-TrFE$_{0.35}$ & 363		& 8 @ 300  & 50 && \cite{furokawa_89}   \\ 
~Phenazine-H$_2$ca & 253 & 1.8 @ 160 & 3$\cdot 10^3$ && \cite{horiuchi_05} \\ 
~Croconic acid & $>$400 & 21 @ 300 & -- && \cite{horiuchi_10} \\ 
~DIPA-Br	& 420 & 10 @ 400 & -- && \cite{fu_13} \\ 
~TTF-BA	& 50 & 0.12 @ 19 & 20 && \cite{kagawa_10} \\ 
~TTF-CA  & 81 & 6.3 @ 59 & 500 && \cite{kobayashi_12} \\ 
		\hline 
\end{tabular}
\begin{tabular}{|l|} 
$^a$ VDF$_{0.65}$-TrFE$_{0.35}$: Vinylidene fluoride-Trifluoroethylene copolymer; H$_2$ca: Chloranilic acid;\\DIPA: Diiso-propylammonium; TTF: Tetrathiafulvalene; BA: Bromanil; CA: Chloranil. \\
	 \hline
\end{tabular}
\label{table}
\end{table*}

The phase transition between paraelectric and ferroelectric phase is related to the microscopic origin of the ferroelectricity. In general, it is a disorder-order type if the paralectric phase is characterized by thermally disordered electric dipoles which on the average cancel out the net polarization, or it is a displacitive transition if the electric dipoles in the ferroelectric phase are originated by a collective shift of the barycenter of the opposite charges. Displacitive transitions are characterized by a soft phonon mode, that is a phonon connected to the charges’ motion, whose frequency goes to zero at the transition. In the proximity of the phase transition the dielectric constant $\kappa$, i.e., the response of the system to the electric field, increases up to several orders of magnitude, and peaks at $T_c$, as shown in the right side of Figure \ref{fig:fig1ferro}.

Table \ref{table} reports the above described basic parameters (the Curie temperature $T_c$, the spontaneous polarization $\boldsymbol{P}_s$ at $T$ close to $T_c$, and the maximum value of the dielectric constant $\kappa$) for some representative ferroelectrics. They have been chosen to represent different classes, both from the point of view of chemical composition (inorganic, organic, organic-inorganic, polymeric, single or double molecular component) and from the microscopic origin of ferroelectricity.

The Rochelle salt, the firstly discovered ferroelectric, is a mixed Na-K tetrahydrated salt of tartaric acid, and the para- to ferro-electric transition is a disorder to order type, implying the position of the ions and possibly of the hydrogen bonds. The shift of the H atom in hydrogen bonded framework is at work also in thiourea, croconic acid, and the co-crystals involving chloranilic acid. On the other hand, the para- to ferro-electric transition of BaTiO$_3$ is a typical example of displacitive transition, as it is for the ms-CT crystals TTF-BA and TTF-CA, with the difference that in the latter cases the displacement involves molecules rather than atoms.

\section{Electronic ferroelectricity}
As Table \ref{table} shows, progress in organic ferroelectrics has been noticeable in the last few years, reaching performance comparable to that of the prototype BaTiO$_3$. What makes ms-CT crystals particularly interesting is that they may display a new type of mechanism for ferroelectricity, which has been named ``electronic'' ferroelectricity to distinguish it from the conventional, or ``ionic'', ferroelectricity \cite{horiuchi_14}. In fact, in conventional ferroelectricity the dipole moments arise from the displacement of oppositely charged atoms/molecules in the crystal, or from the displacement of hydrogens in the hydrogen-bond framework.  Also in the fully ionic ($\rho \sim $1) TTF-BA the ferroelectricity arise from the displacement of the molecular ions TTF and BA when the stack dimerizes due to the Peierls transition. For TTF-CA, on the other hand, the direction of the polarization is opposite to that due to the molecular ions displacement \cite{kobayashi_12}. This unexpected result has been theoretically understood on the basis of modern polarizability theory, showing that the net polarization is the difference between the polarization due to the ion displacement, $\boldsymbol{P}_\mathrm{ion}$, and the larger and opposite polarization due to the shift of the $\pi$-electron cloud within the DA dimer in the chain, $\boldsymbol{P}_\mathrm{el}$ \cite{giovannetti_09}. The electronic response is estimated to be about 20 times as large than the ionic one, and, most importantly, also much faster. All this has been demonstrated for TTF-CA, below the N-I and dimerization transition temperature, namely, below 80 K. After the TTF-CA discovery \cite{kobayashi_12}, the research has been addressed to find ms-CT crystals which exhibit electronic ferroelectricity at higher temperatures, possibly at room temperature. The large amount of experimental data and theoretical modeling available for ms-CT crystals and their peculiar N-I phase transition help to establish the main conditions that have to be fulfilled to attain electronic ferrolectricity.

According to the general requirement for ferroelectricity, inversion center symmetry must be lacking, that for 1:1 ms-CT crystals means that the stack must be dimerized. The dimerization is induced by the Peierls mechanism, namely by response of the system to the electron-phonon coupling. The response is effective for systems with $\rho > $ 0.3, and is maximum in the proximity of the N-I boundary, so in general fully neutral crystals have regular stacks also at low temperatures, intermediate ionicity crystals can have dimerized stacks also at room temperature or in the proximity of it, whereas fully ionic ($\rho >$ 0.8) stacks dimerize at low temperatures. The response of the electronic system to external perturbations is also maximum for systems close to the N-I boundary, so one has to look for such systems to achieve electronic ferroelectricity. As a further requirement, either the unit cell has to contain only a DA pair, or two pair have to present in-phase dimerization. This is the case for TTF-CA, whereas a slightly different system, Dimethyl-tetrathiafulvalene-Chloranil (DMeTTF-CA) presents dimerization and intermediate ionicity, but the two DA dimeric units in the unit cell are arranged out-of-phase, so the system is antiferroelectric \cite{horiuchi_03}.

\subsection{Lock-arm Supramolecular Ordering (LASO) systems}
\begin{figure}[hp]
	\centering
	\includegraphics[width=1.\linewidth]{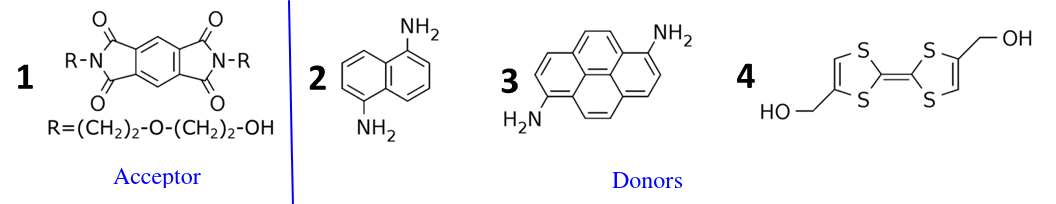}
	\caption{Structural formulas of A and D molecules used in Ref. \cite{tayi_12}.}
	\label{fig:tayimol}
\end{figure}

The goal of room temperature electronic ferroelectricity seemed to have been achieved when Tayi \textit{et al.} \cite{tayi_12} reported ferroelectric hysteresis cycles for three ms-CT crystals obtained on the basis of a novel supramolecular design concept, the Lock-Arm Supramolecular Ordering (LASO), that synergistically combines intermolecular CT and hydrogen bonds. Ferroelectric behavior in three complexes formed by the same electron acceptor \textbf{1} and three different donors, \textbf{2}, \textbf{3} and \textbf{4}. (Figure \ref{fig:tayimol}) was ascribed to a sizeable charge $\rho$ transferred from D to A molecules arranged in non-centrosymmetric structure characterized by dimerized stacks, as TTF-CA below 80 K \cite{masino_17}. A degree of CT of 0.67, 0.89 and 0.43 was estimated through infrared (IR) spectroscopy for co-crystals $\mathbf{1\cdot2}$, $\mathbf{1\cdot 3}$ and $\mathbf{1 \cdot 4}$, respectively. The spontaneous polarization $\boldsymbol{P}_\mathrm{s}$ at room temperature was reported to be about 2 $\mu$C cm$^{-2}$  for all the three co-crystals \cite{tayi_12}.

However, a theoretical calculation by D'Avino and Verstraete \cite{davino_14} suggested that the LASO systems have a much lower degree of CT, hence supporting a very small polarization. At the same time, a re-examination of the X-ray diffraction data of LASO complexes by the proposing team showed that all the three crystal structure are actually better resolved in terms of centro-symmetric space groups, apparently incompatible with ferroelectricity \cite{blackburn_14}. At this point an international collaboration \cite{davino_17} set up to independently reproduce the data of Ref. \cite{tayi_12}. The co-crystal $\mathbf{1\cdot2}$ was selected and synthesized following the original recipe. The X-ray analysis confirmed that the crystal belongs to the triclinic $P\overline{1}$ space group, with two DA pairs per unit cell. Density functional theory (DFT) calculations confirmed that the $P\overline{1}$ phase of
$\mathbf{1\cdot2}$ is the most stable, and that geometry optimization of 
$\mathbf{1\cdot 3}$ and $\mathbf{1 \cdot 4}$ also leads to non-polar structures \cite{davino_17}.

\begin{figure}
	\centering
	\includegraphics[width=1.0\linewidth]{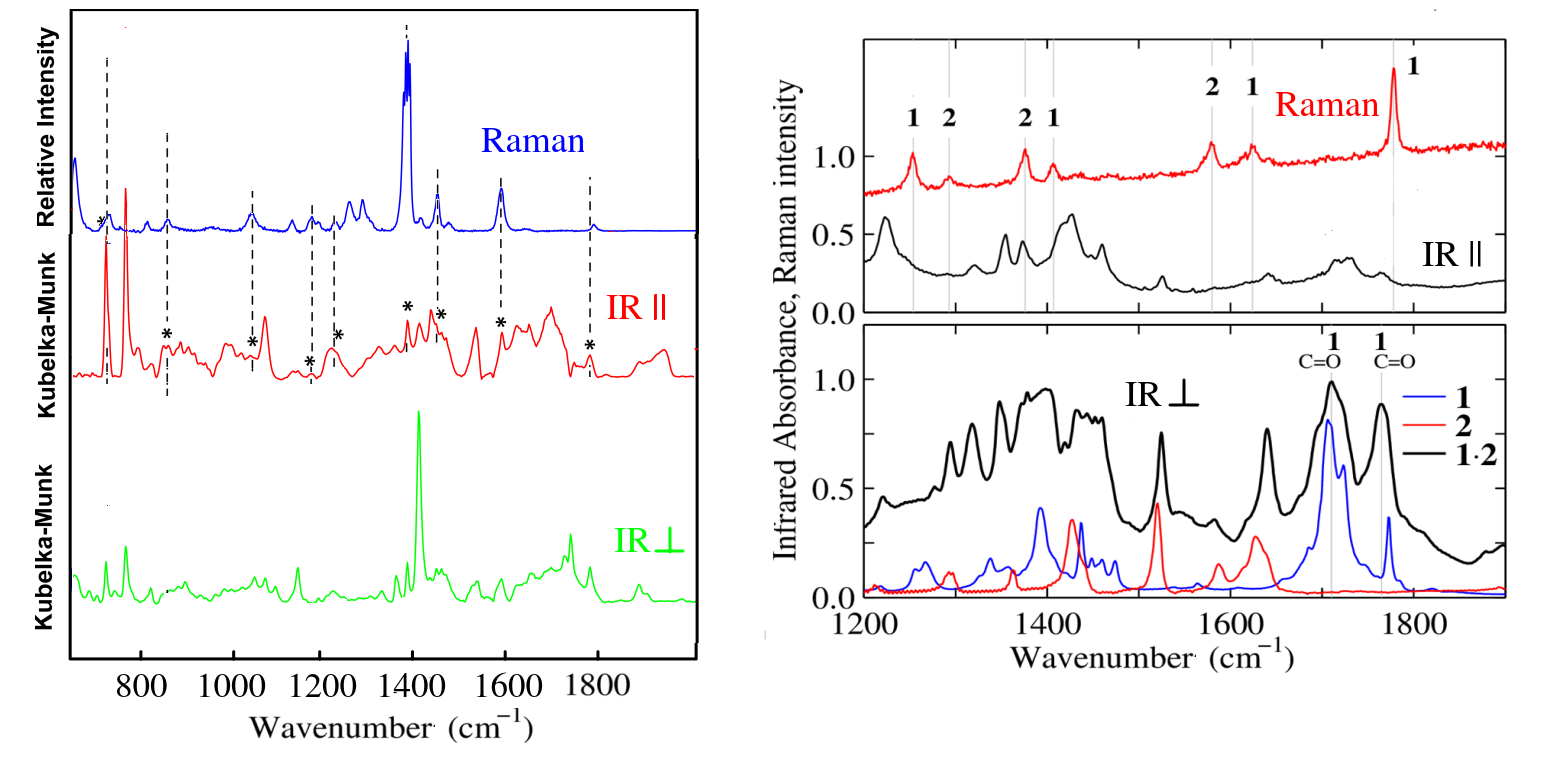}
	\caption{Raman and IR spectra of crystal $\boldsymbol{1\cdot 2}$. Left side, traces from top to bottom: Raman spectrum, IR spectrum polarized along the stack axis and perpendicular to it (adapted from Ref. \cite{tayi_12}); Right side: Top panel, Raman and IR absorption spectrum polarized along the stack axis; bottom panel, IR absorption spectrum polarized perpendicular to the stack axis compared with the spectra of the neutral components \textbf{1} and \textbf{2} (adapted from Ref. \cite{davino_17}).}
	\label{fig:rirtaynoi}
\end{figure}

Further characterization was performed by vibrational spectroscopy. Figure \ref{fig:rirtaynoi} compares the polarized IR and Raman spectra of the co-crystal $\boldsymbol{1\cdot 2}$ obtained in Ref. \cite{tayi_12} (left side) and in Ref. \cite{davino_17} (right side). Vibrational spectra enable to ascertain the presence or absence of an inversion center (regular or dimerized stack), offering complementary information with respect to X-rays. In fact, the latter probes only the long-range order, while vibrational spectroscopy is also sensitive to local dipolar fluctuations and disorder, which might be at the origin of ferroelectricity. Totally symmetric molecular vibrations show up with large or huge intensity in IR spectra only upon dimerization, where they modulate asymmetric flows of electronic charge \cite{masino_17}. The simultaneous presence of these modes in IR and Raman spectra polarized along the DA stack axis is the typical signature of a (possibly local) symmetry breaking. Tayi \textit{et al.} \cite{tayi_12} claimed to have found such a coincidence (asterisks and dashed lines in the left side of Figure \ref{fig:rirtaynoi}), but the frequency coincidences are probably accidental, due to the presence of many bands in their limited quality IR spectra. In fact, IR spectra have been obtained by applying the Kubelka-Munk transformation to ``single point reflectance'' spectra of single crystals, whereas this transformation is only appropriate for the diffuse reflectance spectra of powders. Indeed IR absorbance and Raman spectra by D'Avino \textit{et al.} \cite{davino_17} do not present coincident peaks (Figure \ref{fig:rirtaynoi}, right side), therefore excluding a possible dimerization of the stack.

Infrared spectroscopy allows also the estimation of the ionicity $\rho$ through the frequency shift of properly chosen ``charge sensitive'' vibrational modes. Tayi \textit{et al.} \cite{tayi_12} attributed a sizeable CT to the three LASO compounds on the basis of  analysis conducted on the carbonyl modes of compound \textbf{1}, which appear with polarization perpendicular to the stack (bottom spectrum of the left side of Figure \ref{fig:rirtaynoi}).  Apart from the already mentioned problem of IR spectra quality, Tayi \textit{et al.} \cite{tayi_12} used a rather incomprehensible calibration procedure employing TCNQ complexes, which lead to a frequency increase upon the addition of an electron. As a matter of fact, according to chemical intuition and DFT calculations, the frequency of carbonyl modes is expected to largely decrease upon negatively charging \textbf{1}. Independently from calibration procedures, the bottom panel of the right side of Figure \ref{fig:rirtaynoi} compares the IR spectra of neutral \textbf{1} and \textbf{2} with the one of $\boldsymbol{1 \cdot 2}$ (polarization perpendicular to the stack).  The latter is clearly the superposition of those of its single neutral components in the whole spectral range and specifically in the carbonyl region. This definitely proves that the  $\boldsymbol{1 \cdot 2}$ complex is essentially neutral ($\rho \sim 0$), as the other two LASO compounds are, owing to the minimal shifts of the carbonyl modes. The absence of intermolecular charge transfer in the ground state of the three LASO compound has been also confirmed by DFT calculations that did not evidence any sizeable difference among the electronic structure of the three systems \cite{davino_17}. As a matter of fact, largely neutral systems, more akin to van der Waals crystals rather than CT salts, are not susceptible to dimerization on the basis of the Peierls mechanism \cite{masino_17,davino_17a}.

The crystallographic and spectroscopic data presented by D'Avino \textit{et al.} \cite{davino_17} exclude any indirect signature of electronic ferroelectricity. In addition, electrical polarization hysteresis measurements at various temperatures between 7 and 400 K, with different fields and frequencies, failed to obtain a hysteresis loop, casting doubts on the claimed ferroelectricity of $\boldsymbol{1 \cdot 2}$, and indirectly on that of the other two co-crystals as well \cite{davino_17}. In response to the criticism,   Tayi \textit{et al.} \cite{tayi_17} reproduced the hysteresis loop for $\boldsymbol{1 \cdot 3}$, and reported ferroelectricity for another LASO crystal, made up by \textbf{1} and a variant of
\textbf{2} (one NH$_2$ group substituted by OH) \cite{narayanan_17}. This latter co-crystal, having a 2:1 ratio of A and D and crystallizing in $P\overline{1}$ space group, has been reported to display second harmonic generation (SHG), indicative the lack of inversion center. The authors attribute the discrepancy between crystallographic structure and SHG to hydrogen atoms, not detected by X-ray.

\subsection{Tetramethylbenzidine-Acceptor series}

Collective proton transfer phenomena are a known source of ferroelectricity in molecular systems characterized by the presence of hydrogen bonds \cite{goldsmith_59,horiuchi_05,horiuchi_10}. Apart from the question of reproducibility, this fact may provide an alternative explanation for the ferroelectricity of LASO systems and its elusiveness to X-rays or vibrational spectroscopy. Therefore the quest for room temperature electronic ferroelectricity cannot yet be considered as successful. Another series of ms-CT co-crystal presently being investigated concerns the strong electron donor, 3,3,5,5-Tetramethylbenzidine (TMB) coupled with a series of $\pi$-electron molecules of increasing acceptor strength \cite{castagnetti_18}. The co-crystal of TMB with Tetracyanoquinodimethane (TCNQ) at room temperature has a degree of CT of about 0.3. Around 200 K it undergoes a first-order valence instability transition, with a small increase of $\rho$ to 0.4, as measured by the decrease of the C=C frequency from 1532 to 1528 \cm~ (left side of Figure \ref{fig:tmb-tqf4ir}).   The simultaneous appearance of strong bands in the IR spectra polarized along the stack (right side of Figure \ref{fig:tmb-tqf4ir}) at the same frequency of totally symmetric Raman modes unambiguously signals the dimerization of the stack \cite{castagnetti_17}.
\begin{figure}[t]
	\centering
	\includegraphics[width=1.\linewidth]{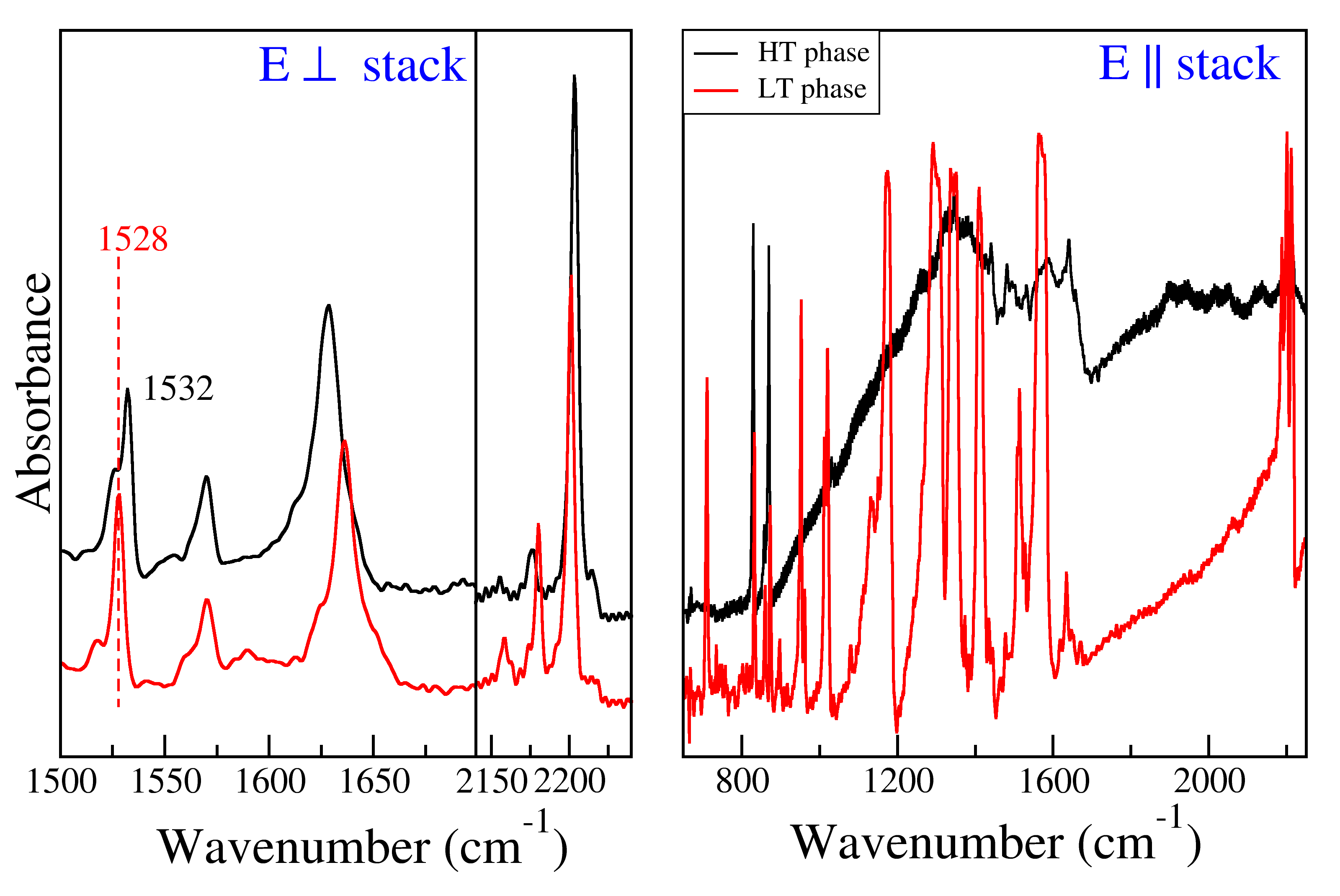}
	\caption{IR spectra of TMB-TCNQ above (HT) and below (LT) the phase transition. Left panel: Spectra polarized perpendicular to the stack. Right panel: Spectra polarized parallel to the stack. From Ref. [23]}
	\label{fig:tmb-tqir}
\end{figure}

\begin{figure}[b]
	\centering
	\includegraphics[width=\linewidth]{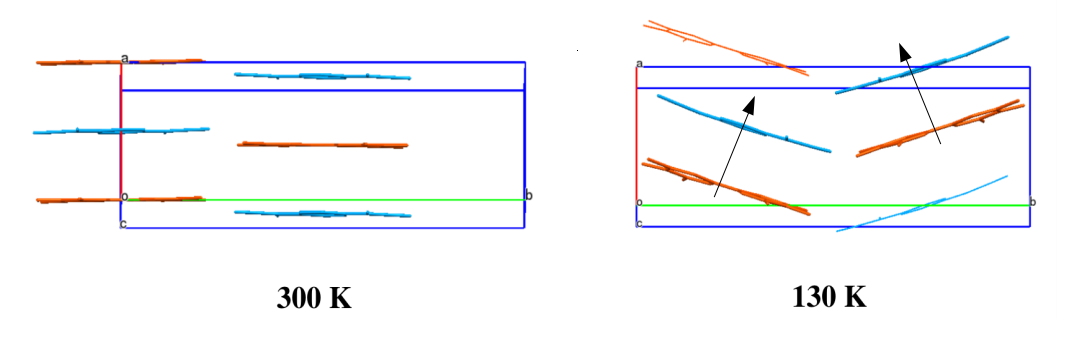}
	\caption{TMB-TCNQ structure at room temperature and 130 K viewed along the [001] direction. Donor and Acceptor molecules are drawn in red and light blue, respectively. Hydrogens are omitted for clarity. The arrows in the 130 K structure evidence the ferroelectric arrangements of the DA dimers.}
	\label{fig:tmb-tqxrayphtr}
\end{figure}

The X-ray structure of the low-temperature phase confirms the loss of the inversion center \cite{mezzadri_18}. The unit cell contains two DA pairs, arranged ferroelectrically (right side of Figure \ref{fig:tmb-tqxrayphtr}). If we compare the valence instability of TMB-TCNQ with the transition of TTF-CA, we see strong analogies. The change of space group at the transition is the same, from $P2_1/n$ to $Pn$, and in both cases the ionicity is close to the N-I interface (0.6 \textit{vs} 0.4). Therefore also TMB-TCNQ might exhibit electronic ferroelectricity like TTF-CA, but at higher temperature (below 200 K rather than below 80 K). Polarization measurements are underway, but without many perspectives of detecting ferroelectricity. First of all, the crystal is damaged at the transition, since it contracts along one crystallographic axis and expands along the other two \cite{mezzadri_18}, likely spoiling the polarization measurements. In addition, the transition implies a strong dimerization, with molecular reorientation (Figure \ref{fig:tmb-tqxrayphtr}) so it is likely that the polarization cannot be easily changed by the electric field: We could have a pyroelectric, polar crystal, but no ferroelectricity.

\begin{figure}
	\centering
	\includegraphics[width=0.8\linewidth]{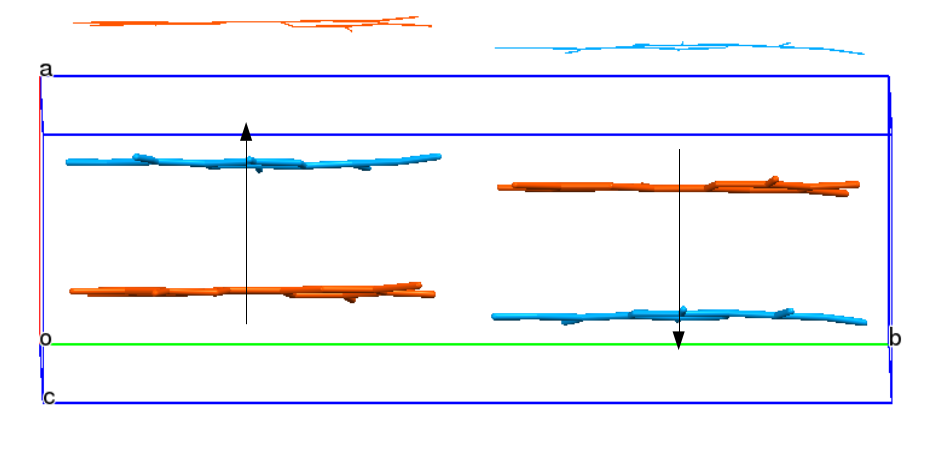}
	\caption{Room temperature crystal structure of TMB-TCNQF$_2$, viewed along the [001] direction. Donor and Acceptor molecules are drawn in red and light blue, respectively. Hydrogens are omitted for clarity. The arrows evidence the anti-ferroelectric arrangements of the DA dimers.}
	\label{fig:tmb-tqf2xray1}
\end{figure}

\begin{figure}
	\centering
	\includegraphics[width=0.8\linewidth]{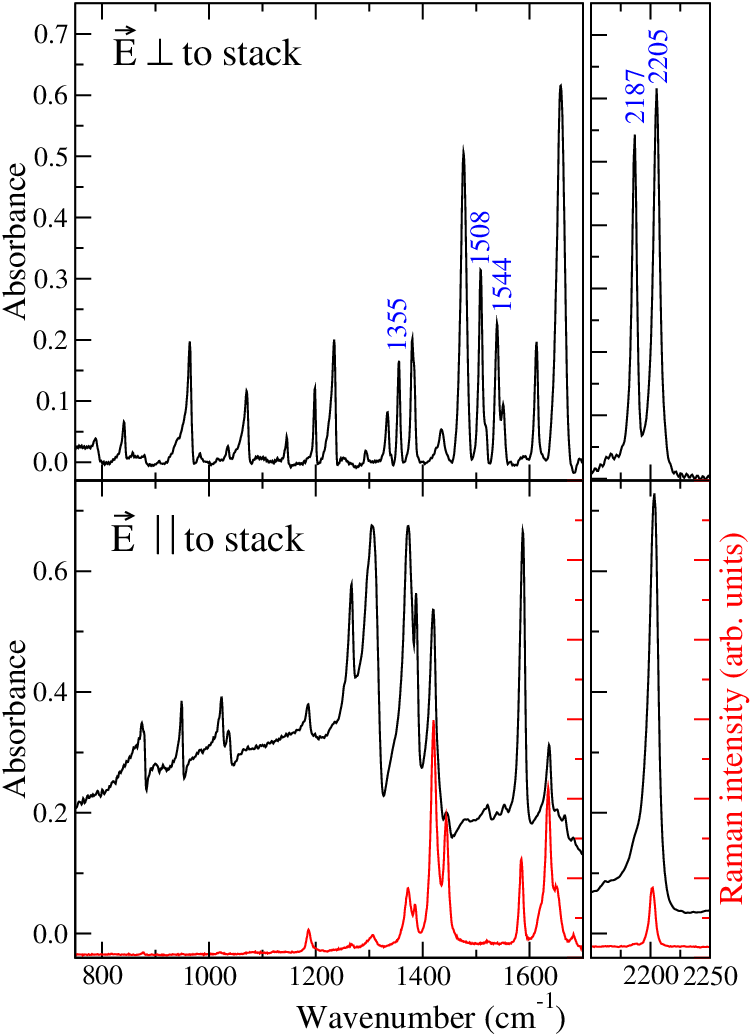}
	\caption{Room temperature polarized IR and Raman spectra of TMB-TCNQF$_4$. From Ref. \cite{castagnetti_18}.}
	\label{fig:tmb-tqf4ir}
\end{figure}

By replacing TCNQ with a stronger electron acceptor, 2,5-Difluoro-TCNQ, the degree of CT of the co-crystal, TMB-TCNQF$_2$, increases from about 0.3 to about 0.7 at room temperature \cite{castagnetti_18}. The stack is dimerized, as shown by IR spectra (not reported here) and by the X-ray analysis. However, the space group ($P2_1$) is different from that of the low-temperature phase of TMB-TCNQ, and the DA dimers are arranged anti-ferroelectrically, as shown in Figure \ref{fig:tmb-tqf2xray1}. 

Higher degree of CT can be attained by combining TMB with Tetrafluoro-TCNQ, TCNQF$_4$ \cite{castagnetti_18}. The identification of TCNQF$_4$ charge sensitive modes, whose frequencies are reported in the upper panel of Figure \ref{fig:tmb-tqf4ir}, yield $\rho$ =  0.9. TMB-TCNQF$_4$ crystallizes in the centrosymmetric monoclinic space group $C2/m$. However, IR spectra polarized along the stack axis compared to the Raman spectrum (Figure \ref{fig:tmb-tqf4ir}, lower panel) show several frequency coincidences, indicating that the inversion center is locally lost. One might suppose that we have a ferroelectric system in the paralectric phase. Measurements are underway, but structural and spectroscopic measurements down to 100 K do not show evidence of phase transitions, retaining the contradiction between X-ray and spectroscopic measurements.

\section{Conclusions}
After about five years, the challenge to find room temperature electronic ferroelectrics is still on the table. As a matter of fact, no ms-CT crystal has been found to exhibit unquestionable electronic ferroelectricity besides TTF-CA, even at low temperature. Some hints about the conditions which have to be met come from the studies reported in this paper. First of all, the ms-CT crystal should have intermediate ionicity, which also implies a strong tendency towards stack dimerization and high response to the electric field, i.e., high polarization. If there are two DA dimers per unit cell, they have to be arranged ferroelectrically  (cf. Figure \ref{fig:tmb-tqxrayphtr}, right side,  and Figure \ref{fig:tmb-tqf2xray1}). Finally, the dimerization should not be too strong or imply molecular distortions, otherwise it might be difficult to reverse, and the system would be pyroelectric, but not ferroelectric.

\end{document}